\documentclass{aa}  
\usepackage{natbib}
\bibpunct{(}{)}{;}{a}{}{,}  
\usepackage{graphicx}
\usepackage{aalongtable}
\usepackage{txfonts}
\begin{document}
   \title{Variability and the X-ray/UV ratio of active galactic nuclei. \\II. Analysis of a low-redshift Swift sample}

   \author{F. Vagnetti \inst{1,2}, M. Antonucci \inst{1}, \and D. Trevese \inst{2}}

   \institute
       {Dipartimento di Fisica, Universit\`a di Roma ``Tor Vergata'', Via della Ricerca Scientifica 1, I-00133, Roma, Italy\\ \email{fausto.vagnetti@roma2.infn.it}
       \and
       Dipartimento di Fisica, Universit\`a di Roma ``La Sapienza'', Piazzale Aldo Moro 2, I-00185 Roma, Italy}   

\date{}

 
  \abstract
   {Variability, both in X-ray and optical/UV, affects the well-known anti-correlation between the $\alpha_{ox}$ spectral index and the UV luminosity of active galactic nuclei, contributing part of the dispersion around the average correlation (``intra-source dispersion''), in addition to the differences among the time-average $\alpha_{ox}$ values from source to source (``inter-source dispersion'').}
   {We want to evaluate the intrinsic $\alpha_{ox}$ variations in individual objects, and their effect on the dispersion of the $\alpha_{ox}-L_{UV}$ anti-correlation.}
   {We use simultaneous UV/X-ray data from Swift observations of a low-redshift sample, to derive the epoch-dependent $\alpha_{ox}(t)$ indices. We correct for the host galaxy contribution by a spectral fit of the optical/UV data. We compute ensemble structure functions to analyse variability of multi-epoch data.}
   {We find a strong ``intrinsic $\alpha_{ox}$ variability'', which makes an important contribution ($\sim40\%$ of the total variance) to the dispersion of the $\alpha_{ox}-L_{UV}$ anti-correlation (``intra-source dispersion''). The strong X-ray variability and weaker UV variability of this sample are comparable to other samples of low-$z$ AGNs, and are neither due to the high fraction of strongly variable NLS1s, nor to dilution of the optical variability by the host galaxies. Dilution affects instead the slope of the anti-correlation, which steepens, once corrected, becoming similar to higher luminosity sources. The structure function of $\alpha_{ox}$ increases with the time lag up to $\sim$1 month. This indicates the important contribution of the intermediate-long timescale variations, possibly generated in the outer parts of the accretion disk.}
   {}

\authorrunning{F.Vagnetti et al.}
\titlerunning{X-ray/UV ratio of AGNs. II}

   \maketitle

\section{Introduction}
The X-ray to UV ratio of active galactic nuclei (AGN) gives direct information on an important  region of the spectral energy distribution (SED), relating the radiative processes operating in the accretion disk and in the corona, connecting their emissions across the unobservable band of the extreme UV. It characterises the shape of the SED, and affects, through the ionisation equilibrium, properties of the UV spectral lines, such as the equivalent width and the blue-shift of the CIV $\lambda$1549 emission line \citep{rich11}.

The X-ray/UV ratio is often expressed through the inter-band spectral index 
\begin{equation}
\alpha_{ox}\equiv{\log(L_X/L_{UV})\over\log(\nu_X/\nu_{UV})}=0.384 \log\left({L_X\over L_{UV}}\right)
\end{equation}
between the conventional frequencies $\nu_X\equiv\nu_{\rm 2keV}$ and $\nu_{UV}\equiv\nu_{2500\AA}$.

$\alpha_{ox}$ is found to be strongly anti-correlated with the ultraviolet specific luminosity $L_{UV}$, showing that more luminous objects are, on average, relatively weaker in X-rays:
\begin{equation}
\alpha_{ox} =a\log L_{UV}+{\rm const}\quad (a<0)\,.
\end{equation}
This relation has been studied by many authors, who found slopes approximately in the interval
$-0.2\la a\la -0.1$ \citep[e.g.,][]{stra05,stef06,just07,gibs08,grup10,vagn10}, depending on the selection of the sample, and especially on its range of luminosities and/or redshifts. For instance, while \citet{just07} gets $a=-0.14$  within a wide area of the $L-z$ plane, $27.5<\log L_{UV}<33$, $0<z<6$, a flatter slope is found by \citet{grup10}, $a=-0.114$, for a low-luminosity and low-redshift sample, $26<\log L_{UV}<31$, $z<0.35$, and a steeper slope, $a=-0.217$, is obtained by \citet{gibs08} for higher redshifts and luminosities, $30.2<\log L_{UV}<31.8$, $1.7<z<2.7$. A similar trend is found dividing a wider sample in two subsamples with lower and higher luminosity or redshift, see e.g. \citet{stef06,vagn10}. Thus, a precise estimate of the slope cannot be done in general terms, as the $\alpha_{ox}-L_{UV}$ relation itself might be non-linear.

Moreover, \citet{gibs08} noticed the large scatter of the data around the average relation, suggesting that a large fraction of it can be due to  variability, combined with non-simultaneity of the X-ray and optical observations. In a previous paper
 \citep[][paper I]{vagn10}, we have analysed a sample with simultaneous measurements extracted from the XMM-Newton Serendipitous Source Catalogue \citep{wats09}, concluding that ``artificial $\alpha_{ox}$ variability'' due to non-simultaneity is not the main cause of dispersion, while ``intrinsic $\alpha_{ox}$ variability'' of individual sources (or ``intra-source dispersion'') and intrinsic differences in the time-average values of $\alpha_{ox}$ from source to source (or ``inter-source dispersion'') are the most important contributions.

In paper I, we then analysed $\alpha_{ox}$ variability computing the ensemble structure function, and pointed out the need of further AGN samples with simultaneous measurements to make progress in this topic. Appropriate data can be obtained by space observatories having both X-ray and optical/UV telescopes onboard, such as {\it XMM-Newton} and {\it Swift}.

In this paper, we present the analysis of a sample of low-redshift AGNs observed by {\it Swift} and previously studied by \citet{grup10}, whose paper and sample will hereafter be referred as G10.

This paper is organised as follows. Section 2 describes the data extracted from G10 and from the {\it Swift} archive. Section 3 analyses the $\alpha_{ox}-L_{UV}$ anti-correlation and its dispersion. In Sect. 4, we present the ensemble structure function of the intrinsic X/UV variability. Section 5 discusses and summarises the results.

Throughout the paper, we adopt the cosmology H$_{o}$=70 km s$^{-1}$ Mpc$^{-1}$, $\Omega_{m}$=0.3, and $\Omega_{\Lambda}$=0.7.
\bigskip\noindent

\section{The data}

The G10 sample consists of 92 AGNs extracted from the bright soft X-ray selected  sample of \citet{grup01}, and observed with {\it Swift} between 2005 and 2010. The sample by \citet{grup01} contains all the 110 Seyferts from the sample of 397 sources by \citet{thom98}, which was extracted from the ROSAT All Sky Survey \citep{voge99} to include sources selected to be X-ray bright (count rate $>$0.5 counts/s), X-ray soft  (hardness ratio\footnote{with reference to the ROSAT bands 0.1-0.4 keV and 0.5-2.4 keV} HR$<$0.0), and at high Galactic latitude ($|b|>20^\circ$).

The G10 sample includes simultaneous X-ray and optical/UV measurements for most of the sources, and in many cases multi-epoch observations are available, with a total of 299 observations for the 92 sources. However, in a few cases the data are not usable for our purposes, because of lack of X-ray or optical/UV measurements. We therefore adopt a preliminary subsample of 90 sources, with 74 multi-epoch sources and 16 single-epoch sources, for a total of 241 observations. 
In the following analysis (Section 2.1), we will remove a few observations where determination of the AGN luminosity is unreliable, because of strong dominance of the host galaxy, or because of insufficient spectral coverage of the optical/UV data. We will define the resulting set of 216 observations as sample A, including 86 sources (68 multi-epoch and 18 single-epoch). In Section 3, we will introduce a further subsample not containing known radio-loud sources, which will be called sample B. The data are taken from the tables of the G10 paper, available electronically, and checked through the Swift archive at Heasarc\footnote{http://heasarc.gsfc.nasa.gov/}.

Compared to the sample of Paper I, the sample studied in the present paper lies in a region of the luminosity-redshift plane at lower redshifts ($z<0.35$) and luminosities ($26 < \log L_{UV} < 31$), see Figure 1. The relevant properties of the sources of samples A and B at each epoch are reported in Table 1, where: Col. 1 corresponds to the source serial number according to G10; Col. 2, source name; Col. 3, observation epoch serial number; Col. 4, epoch, in modified Julian days (MJD); Col. 5, redshift; Col. 6, soft X-ray spectral index, according to Table 4 of G10; Col. 7, logarithm of the specific luminosity at 2 keV in erg s$^{-1}$Hz$^{-1}$; Col. 8, logarithm of the specific AGN luminosity at 2500\AA\ in erg s$^{-1}$Hz$^{-1}$; Col. 9, logarithm of the specific host galaxy luminosity at 2500\AA\ in erg s$^{-1}$Hz$^{-1}$ (substituted by a hyphen when the galaxy contribution is negligible); Col. 10, optical/X-ray spectral index; Col. 11, radio-loudness flag $f_{RL}=1$ (radio-loud),  $f_{RL}=0$ (radio-quiet),  $f_{RL}=-1$ (unclassified).

\begin{figure}
\centering
\resizebox{\hsize}{!}{\includegraphics{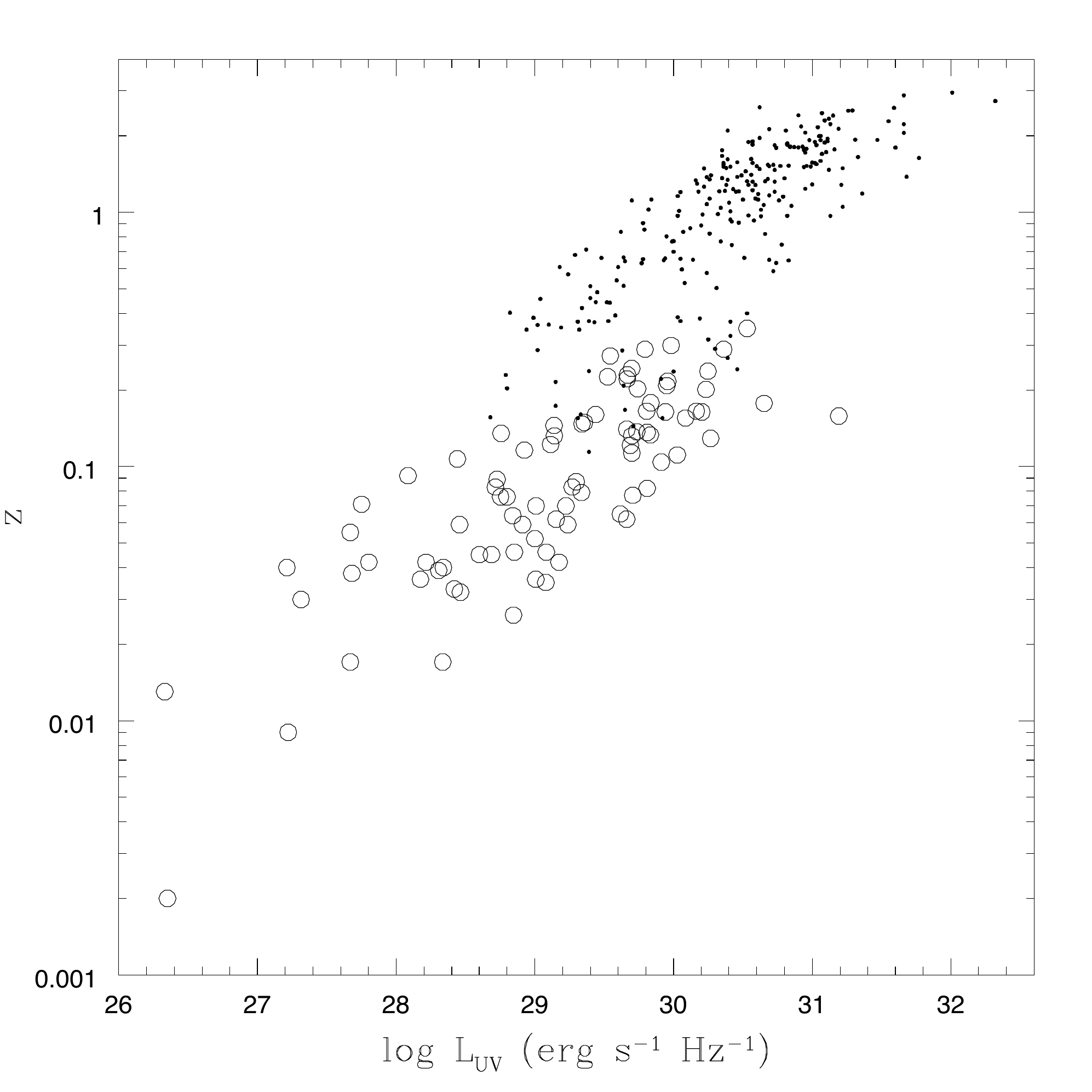}}
\caption{Distribution of the sources in the $L_{UV}$-$z$ plane. Open circles: {\it Swift} sample; dots: {\it XMM-Newton} sample.}
\end{figure}

\subsection{UV luminosities}
The UltraViolet/Optical Telescope (UVOT) onboard {\it Swift} has six photometric filters, whose central wavelengths are, respectively, $\lambda(V)=5468$\AA, $\lambda(B)=4392$\AA, $\lambda(U)=3465$\AA, $\lambda(UVW1)=2600$\AA, $\lambda(UVM2)=2246$\AA, and $\lambda(UVW2)=1928$\AA\ \citep{romi05,romi09}. Magnitudes in one or more of these bands are available for each source and epoch of our sample from Table 3 of G10. We first transformed magnitudes to fluxes according to the formulae given by \citet{pool08}, using the count rate to flux conversion factors of their table 10 (GRB models, also appropriate for AGNs). 
To estimate UV luminosities at 2500\AA, similarly to the procedure used in paper I, we compute, from each of the available fluxes, the corresponding luminosities as $L_\nu(\nu_{rest})=F_\nu(\nu_{obs})\, 4\pi D_L^2/( 1+z)$, and derive the rest-frame SEDs, which are shown in Figure 2.


Then, we take into account the contribution of the host galaxy starlight, which can be important for AGNs of low luminosity such as those considered here. Following a procedure similar to that adopted by \citet{luss10}, we model the optical spectrum by a combination of AGN and galaxy components, as

\begin{equation}
L_\nu=A\left[f_AF_R(\nu)+f_G(\nu/\nu_*)^{-3}\right]
\end{equation}
where $F_R(\nu)$ is the mean SED computed by \citet{rich06} for type 1 quasars from the Sloan Digital Sky Survey (SDSS), $A$ is a normalization factor, and the coefficients $f_G$ and $f_A$ represent the galaxy and AGN fractional contributions at the frequency $\nu_*$ corresponding to 2500\AA\ ($\log\nu_*=15.08$). The average spectral index $\alpha_{opt}$ of Eq. (3) is a monotonic function of the ratio $f_G/f_A$, which is thus determined by comparison with the slope of each observed SED. A clear sign of this dependence is apparent in Figure 2, where less luminous sources have progressively steeper spectra.

The normalization factor $A$ is then fitted to the data by general linear least squares \citep{pres92}, as
\begin{equation}
A={\sum_{i=1}^N{y_iX(\nu_i)/\sigma_i^2}\over\sum_{i=1}^N{[X(\nu_i)]^2/\sigma_i^2}}
\end{equation}
where $X(\nu_i)=\log L_\nu(\nu_i)$ is given by the model function of Eq. (3) computed in correspondence of the available UVOT rest-frame frequencies $\nu_i$, $y_i=\log L_i$ is given by the corresponding measured specific luminosities and $\sigma_i$ are their errors. This procedure determines, for each source and epoch, the luminosities of the two components at 2500\AA, $L_{AGN}$ and $L_G$. In most cases, we can compare the different determinations of $L_G$ for the same source at more epochs, finding small dispersions (usually $\la 0.15$ in $\log L_G$). We then fix $L_G$ to its average value, for each source, and repeat the fit to the data modifying the fitting function as
\begin{equation}
L_\nu=A'F_R(\nu)+L_G(\nu/\nu_*)^{-3}\quad,
\end{equation}
where the factor $A'$ is now given by
\begin{equation}
A'={\sum_{i=1}^N{y'_iX'(\nu_i)/\sigma_i^2}\over\sum_{i=1}^N{[X'(\nu_i)]^2/\sigma_i^2}}\quad,
\end{equation}
with $X'(\nu_i)=\log F_R(\nu_i)$, and $y'_i=\log [L_i-L_G(\nu_i/\nu_*)^{-3}]$. The AGN luminosity at 2500\AA\ is then given by $L_{AGN}=A'F_R(\nu_*)$. In the following, we refer to it simply with $L_{UV}$, maintaining the name $L_G$ for the galactic contribution.

For a subsample of 10 sources, {\it Hubble Space Telescope} observations by \citet{bent09} are available, with direct measurements of the AGN and galactic luminosities. Our estimated values of $L_G$ are consistent with such measurements.

In some cases, when the number of available UVOT data is $< 4$, only a small portion of the SED is sampled, and the two contributions cannot be determined. This occurs for 19 observations in total, leading to the removal of 2 sources from our sample, and to the decrease of the number of useful observations for some of the remaining sources. We also remove 2 more sources, for which the slope of the observed SED is steeper than -3, indicating a negligible AGN contribution. Therefore, we remove 4 sources in total, so defining our sample A, which includes 86 sources (68 multi-epoch and 18 single-epoch), for a total of 216 observations.  The galactic dilution is substantial ($f_G\ga 30\%$) for 15 sources out of the 86 sources of sample A. This will be further discussed in Section 3.3.

\begin{figure}
\centering
\resizebox{\hsize}{!}{\includegraphics{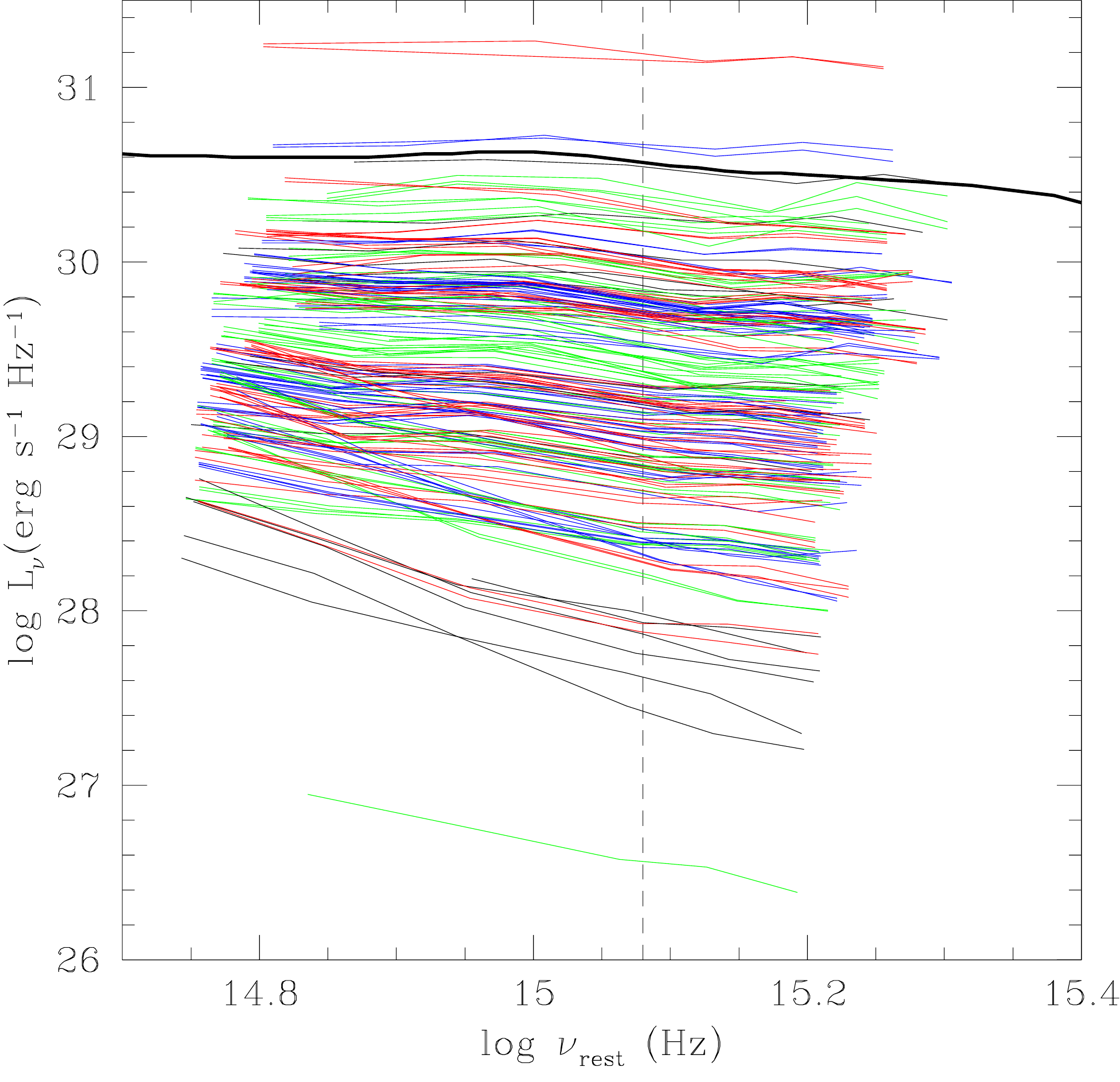}}
\caption{Spectral energy distributions from the available UVOT data. Black lines refer to sources with data at a single epoch, while coloured lines refer to multi-epoch sources. Data from the same source are plotted with the same colour, but more sources are represented with the same colour. The continuous curve covering all the range of the plot is the average SED computed by \citet{rich06} for type 1 quasars from the SDSS.}
\end{figure}

\subsection{X-ray luminosities}
Unabsorbed rest-frame soft-band X-ray fluxes, $F_X({\rm 0.2-2\,keV})$, are given by G10, together with the soft-X-ray spectral index $\alpha_x$ (defined according to the rule $F_\nu\propto\nu^{-\alpha_x}$). We derive the specific flux at 2keV as

\begin{equation}
F_\nu({\rm 2\,keV})={F_X({\rm 0.2-2\,keV})\over \nu_{\rm 2\,keV}} f(\alpha_x)\, ,
\end{equation}
with $f(\alpha_x)={(\alpha_x-1)/(10^{\alpha_x-1}-1)}$ for $(\alpha_x\ne 1)$ and $f(\alpha_x)={1/\ln10}$ for $(\alpha_x=1)$, and compute the specific luminosities accordingly.

\section{The $\alpha_{ox}-L_{UV}$ anti-correlation}
Radio-loud (RL) quasars are known to be relatively X-ray bright because of the enhanced X-ray emission associated with their jets \citep[e.g.][]{zamo81,worr87}; in contrast, broad absorption line (BAL) quasars are relatively X-ray faint, compared to non-BAL quasars \citep[e.g.][]{gree96,bran00,gibs08}. Both populations are therefore usually removed in the analysis of the $\alpha_{ox}-L_{UV}$ anti-correlation \citep[e.g.][]{just07,gibs08,youn10,vagn10}.

We find radio information from the NASA/IPAC Extragalactic Database (NED) for 35 sources out of 86; we use the data at 5 GHz when available, or scale the flux as $f_\nu\propto\nu^{-0.8}$ when observations are available at a different frequency. We classify the sources as RL when the inequality $R^*=L_\nu(5{\rm\,GHz})/L_{UV}>10$ is satisfied \citep[e.g.,][]{sram80,kell89}, marking them with $f_{RL}=1$ in Table 1 (9 sources out of 35). Sources with $R^*< 10$ are classified as radio-quiet (RQ) and marked with $f_{RL}=0$ (26 out of 35). Sources without radio information (51 out of 86) are marked with $f_{RL}=-1$. Concerning the presence of BAL quasars among our sources, we have checked a number of studies about low-redshift BALs \citep{pett85,turn86,kinn91,turn97,sule06,gang07}, finding no coincidences. Although both radio and BAL information are quite incomplete, we finally remove only 9 RL AGNs from sample A, so defining a reference sample of 77 sources (61 multi-epoch + 16 single epoch, sample B) for our subsequent analysis. Sample B includes 194 observations listed in Table 1 with $f_{RL}\neq 1$.

Figure 3 shows the distribution of sources (sample A, circles and triangles; sample B, circles) in the plane $\alpha_{ox}-L_{UV}$, compared with the {\it XMM-Newton} sample (dots) studied in Paper I. The average values of $\alpha_{ox}$ and $L_{UV}$ are shown for multi-epoch sources. Also shown is the linear least squares fit for the {\it Swift} sample B:

\begin{equation}
\alpha_{ox}=(-0.135\pm0.015)\log L_{UV}+(2.645\pm0.446)
\end{equation}
(thick continuous line). Moreover, the fit for the {\it XMM-Newton} reference sample of Paper I, $\alpha_{ox}=(-0.178\pm0.014)\log L_{UV}+(3.854\pm0.420)$ (thin continuous line), and the fit obtained by G10, $\alpha_{ox} =(-0.114 \pm 0.014) \log L_{UV} + (1.975 \pm 0.403)$ (with luminosities scaled to cgs units, dotted line) are shown. Our present fit is somewhat steeper than that of G10, due to the correction that we operate for galactic dilution. Both fits are much flatter than our fit of Paper I, which is derived from higher luminosity sources. For further comparison, the fits by \citet{just07}, $\alpha_{ox} =(-0.140 \pm 0.007) \log L_{UV} + (2.705 \pm 0.212)$, and by \citet{gibs08}, $\alpha_{ox} =(-0.217 \pm 0.036) \log L_{UV} + (5.075 \pm 1.118)$ are shown. There is a clear tendency for a steepening of the $\alpha_{ox}-L_{UV}$ anti-correlation for samples extending at higher luminosities, as already mentioned in the introduction, and discussed in previous works \citep{stef06,vagn10}.

\begin{figure}
\centering
\resizebox{\hsize}{!}{\includegraphics{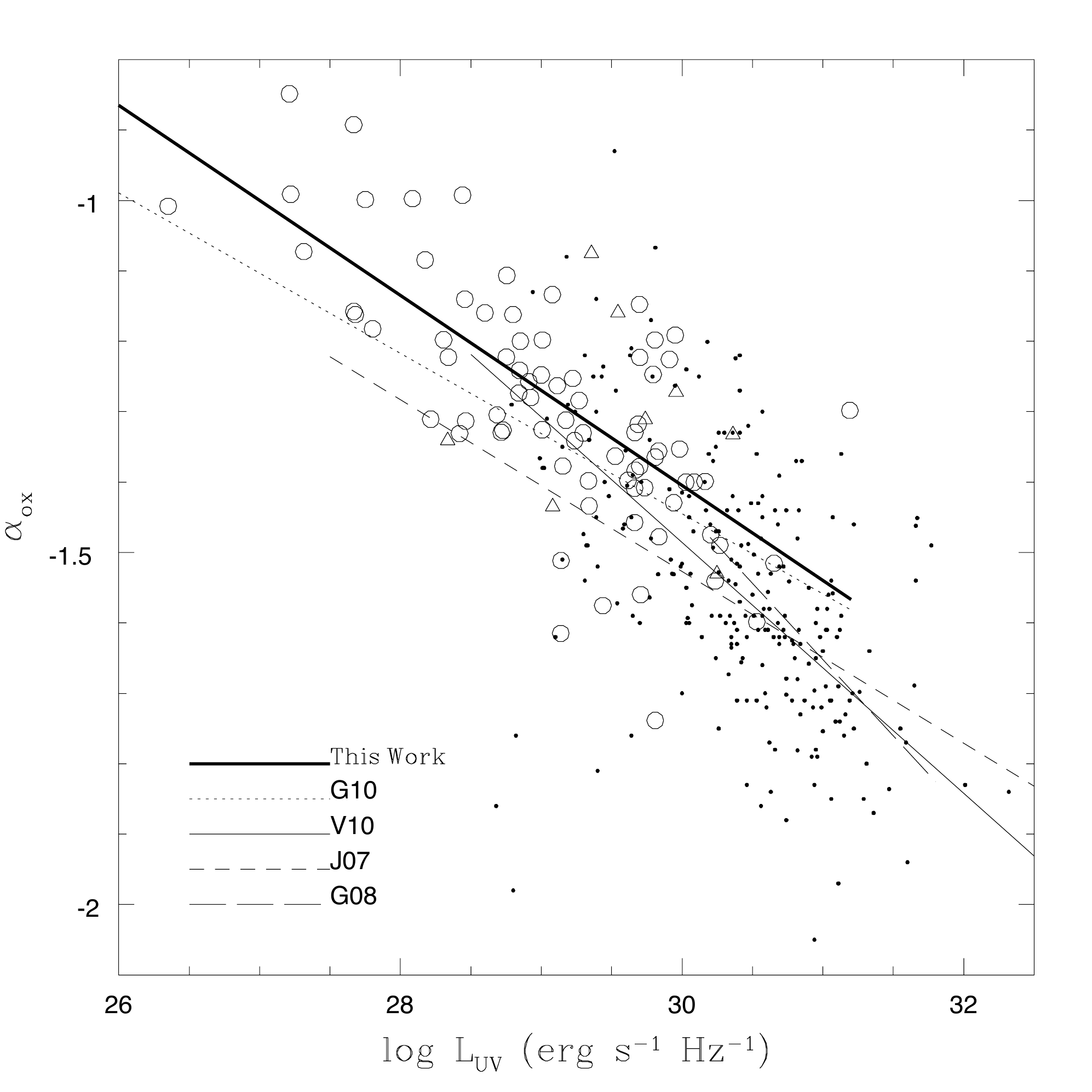}}
\caption{$\alpha_{ox}$ as a function of the 2500\AA\ specific luminosity $L_{UV}$, for samples A (circles and triangles) and B (circles) and for the sample of Paper I (dots). Triangles refer to radio-loud sources, circles to radio-quiet and radio-unclassified sources. Linear fits are shown for the present work, and for previous works marked G10 \citep{grup10}, V10 (Paper I), J07 \citep{just07}, G08 \citep{gibs08}.}
\end{figure}

\subsection{Dispersion}

We define the residuals
\begin{equation}
\Delta\alpha_{ox}=\alpha_{ox}-\alpha_{ox}(L_{UV})\quad ,
\end{equation}
adopting Eq. (8) as our reference $\alpha_{ox}(L_{UV})$ relation. The standard deviation of our distribution of the residuals is $\sigma=0.124$ for sample A, and $\sigma=0.117$ for sample B.
The dispersion in our $\Delta\alpha_{ox}$ distribution is of the same order as those obtained in some studies based on non simultaneous X-ray and UV data. Indeed, it is smaller than those found by \citet[][e.g.]{stra05} (0.14) and by \citet{youn10} (0.16), but slightly larger than that evaluated by \citet{gibs08} (0.10). Values found in previous simultaneous studies are also of the same order, e.g. our {\it XMM-Newton} sample of Paper I (0.12), and the small clean catalog by \citet{wu12} (0.12). So we confirm our conclusion of Paper I, that non-simultaneity of X-ray and UV measurements, that we call ``artificial $\alpha_{ox}$ variability'', is not the main contribution to the dispersion of the residuals $\Delta\alpha_{ox}$. \citet{wu12} reach the opposite conclusion, but they compare their results only with those of \citet{just07} (0.15). Non-simultaneity would lead to an ``artificial'' change of $\alpha_{ox}$ caused by the sole change of the X-ray flux in the time elapsed from the optical measurement, or viceversa. An average change of 15-30\% in a few years would apply for the optical case \citep[see e.g.][]{wilh08,macl12}, and 40-50\% for the X-ray case \citep{mark04,vagn11}. Applying Eq. (1), this would translate to an $\alpha_{ox}$ ``artificial'' change of $\la 0.07$. On the other hand, as we have shown in Paper I, and as we will discuss further below, there is a sizable ``intrinsic $\alpha_{ox}$ variability'' which we estimated $\sim 0.07$, large enough to provide a significant contribution, at least of the same order as artificial variability, even when the latter is removed by simultaneous X/UV measurements.

\begin{figure}
\centering
\resizebox{\hsize}{!}{\includegraphics{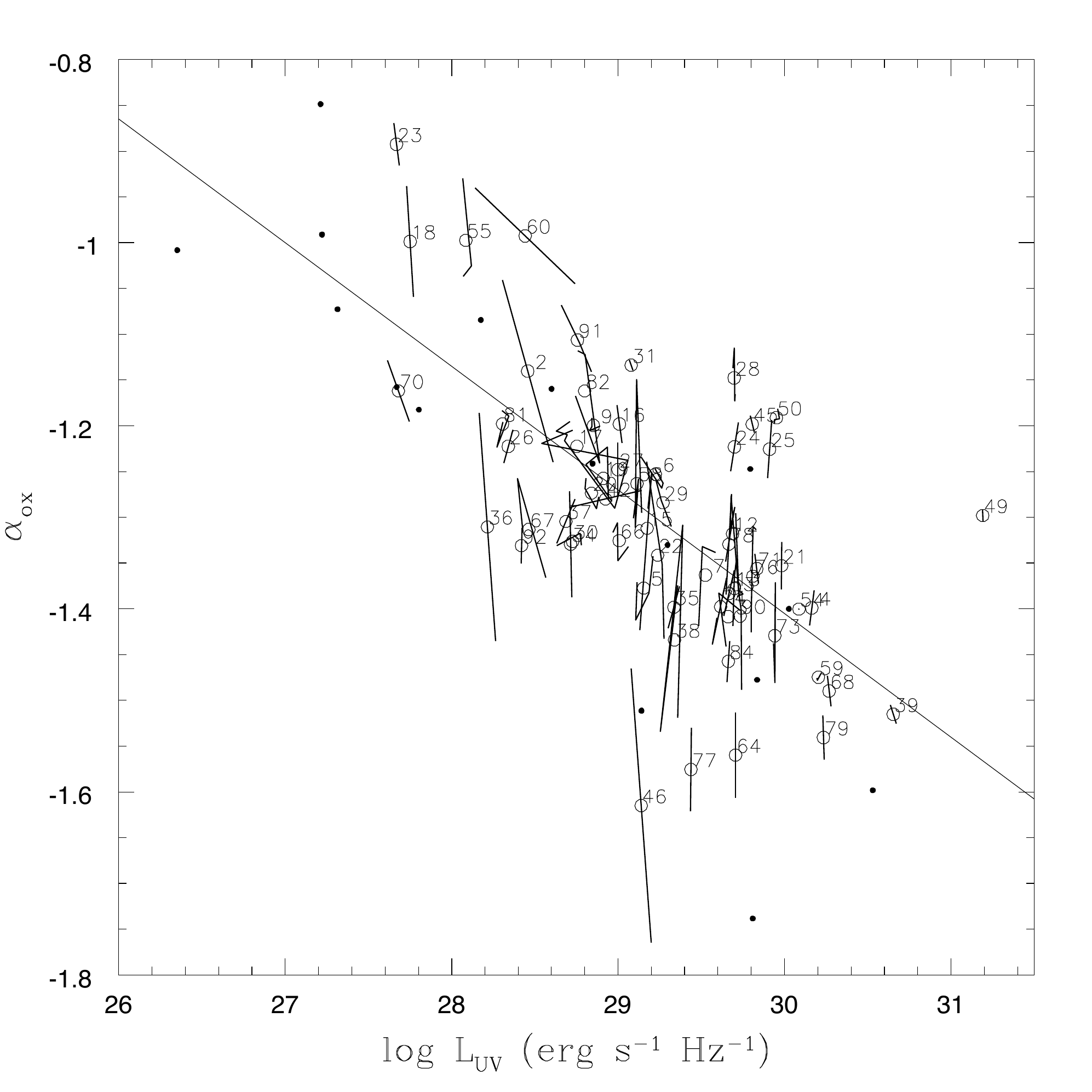}}
\caption{Tracks of individual sources of sample B in the plane $\alpha_{ox}-L_{UV}$. Connected segments show the tracks of multi-epoch sources, while open circles represent the average values of the same sources, which are labeled with their serial numbers as in Table 1. Objects with single-epoch measurements are represented by dots. The straight line is the adopted $\alpha_{ox}-L_{UV}$ relation, Eq. (8).}
\end{figure}

\subsection{Tracks of individual sources}

Multi-epoch information is available for 68/86 sources of sample A, and for 61/77 sources of sample B. We show in Figure 4 the tracks of individual sources in the $\alpha_{ox}-L_{UV}$ plane, for sample B. Large variations in $\alpha_{ox}$ are clearly occurring for many sources, and most tracks appear almost vertical, suggesting the occurrence of strong changes in X-rays, and/or weak changes in $L_{UV}$. This is confirmed by the histograms of the individual variability dispersions of $\log L_{UV}$ and $\log L_X$ for multi-epoch sources, shown in Figure 5. The variations occur on various time scales from days to years, so a better evaluation of the variability properties will be made in the next Section.
We note, however, that some factors could affect this apparent behavior, e.g. the presence of a large number of Narrow Line Seyfert 1 (NLS1) nuclei, which are known to have strong X-ray variability \citep[e.g.][]{leig99} and moderate optical variability \citep[e.g.][]{ai10}. We have instead corrected for the effect of the other important factor, dilution of the optical variability by the host galaxy.

\begin{figure}
\centering
\resizebox{\hsize}{!}{\includegraphics{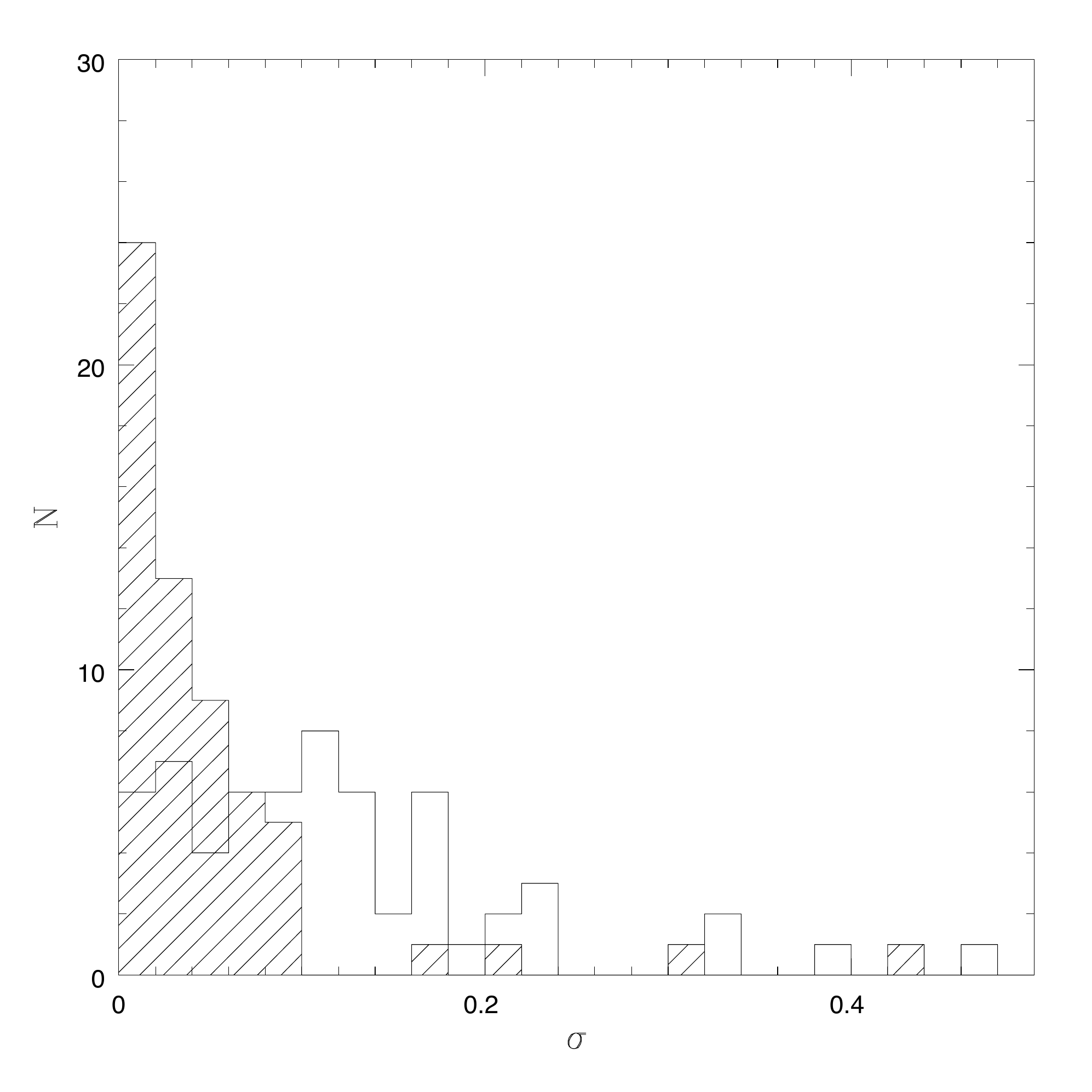}}
\caption{Histograms of the individual variability dispersions of $\log L_{UV}$ and $\log L_X$ for multi-epoch sources: shaded histogram, UV; empty histogram, X-ray.}
\end{figure}

\subsection{Effect of the host galaxy}

Although we have subtracted host galaxy luminosities in Section 2.1, it is useful to discuss the possible effects of such contributions, for comparison with the literature. \citet{wilk94} first pointed out that contamination by host galaxy starlight could affect the $\alpha_{ox}-L_{UV}$ relation, and that excluding the lowest luminosity AGNs would cause a marginal steepening of the relation. G10 mention the possibility that the measured magnitudes are affected by a contribution of the host galaxy starlight within the UVOT standard extraction radius of 5 arcsec, estimating this effect important for a few extreme cases like Mark 493. \citet{wu12} analyse a large sample of quasars on wide $L$ and $z$ intervals, and point out that their $\alpha_{ox}-L_{UV}$ slope decreases from $-0.16$ to $-0.14$ when the G10 sample is added, arguing that the difference in slopes is likely caused by host galaxy contamination at low redshift. \citet{luss10} model the optical spectrum as a combination of AGN and  galaxy components, $L_\nu=A\nu^{-0.5}+G\nu^{-3}$, and estimate the galaxy contribution from the measure of the optical spectral index. This enables the authors to correct their $\alpha_{ox}-L_{UV}$ relation, which results in a steepening from $-0.154$ to $-0.197$. \citet{xu11} analyses a sample of low-luminosity AGNs,  including 28 local Seyfert galaxies and 21 low-ionization nuclear emission-line regions (LINERs), with $L_{UV}$ luminosities in the range $10^{22}-10^{27.7}$ erg s$^{-1}$Hz$^{-1}$. The author takes the nuclear magnitudes directly observed by \citet{ho01} with {\it Hubble Space Telescope}, or estimated from ${H_\beta}$ luminosity, and finds for the relation $\alpha_{ox}-L_{UV}$ a steeper slope ($-0.134$) than G10, but similar to our result of Eq. (8) and to that found by \citet{just07} for higher luminosity AGNs.

While Eq. (8) is corrected for galactic dilution, we have computed the same relation for uncorrected (diluted) luminosities as well, $\alpha_{ox}=(-0.103\pm0.016)\log L_{UV}+ (1.679\pm0.472)$, which is flatter. The dilution effect is shown in Figure 6, where some of the low luminosity sources are shifted towards higher $L_{UV}$ and lower $\alpha_{ox}$, along lines with slope $-0.384$, according to Eq. (1). This slope is higher than the anti-correlation slope, especially in the low-luminosity range, so determining a flattening of the observed anti-correlation.

\begin{figure}
\centering
\resizebox{\hsize}{!}{\includegraphics{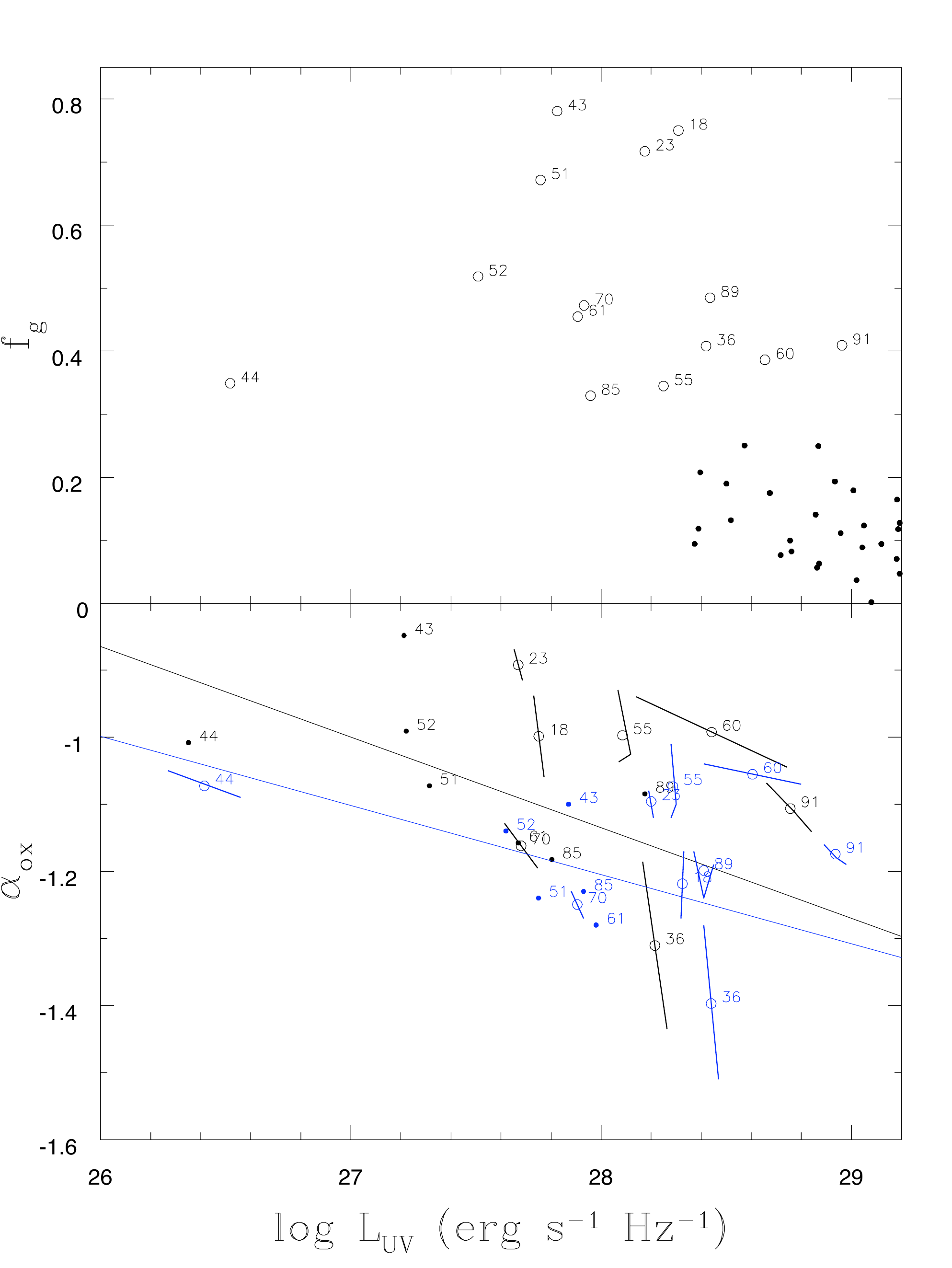}}
\caption{Effect of the dilution by the host galaxy. {\it Upper panel}: galactic fraction $f_g$ as a function of the UV luminosity; sources with $f_g>30\%$ are represented by circles and numbered, sources with $f_g<30\%$ are shown as dots. {\it Lower panel}: the $\alpha_{ox}-L_{UV}$ relation before (blue) and after (black) correction for galaxy dilution. Only the most diluted sources are shown, all shifted along lines with slope -0.384, by amounts increasing with $f_g$. Sources 44 and 89 have a reduced number of epochs after correction, due to the requirement that the SED contains at least 4 UVOT data, as discussed in Section 2.1.}
\end{figure}


\section{The structure functions}

We now compute, for the 59 multi-epoch sources of sample B, an ensemble structure function (SF) to describe the variability of $\alpha_{ox}$ as a function of the rest-frame time lag $\tau$. We define it as in \citet{dicl96}, and in agreement with the procedure used in Paper I:

\begin{equation}
S\negthinspace F_\alpha(\tau)=\sqrt{\pi/2}\,\langle|\alpha_{ox}(t+\tau)-\alpha_{ox}(t)|\rangle\, ,
\end{equation}
where $t$ and $t+\tau$ are two epochs, in the rest-frame, at which $\alpha_{ox}$ is determined. The factor $\sqrt{\pi/2}$ is introduced\footnote{Due to a misprint, an incorrect factor $\pi/2$ was written in Paper I. The correct factor $\sqrt{\pi/2}$ was however used in the computations.} to normalise the SF to the r.m.s. value in the case of a Gaussian distribution, and  the angular brackets indicate the ensemble average over appropriate bins of time lag.

The SF, displayed in Figure 7, shows an average increasing behavior. Maximum variations are $\sim 0.073$ at $\sim 1$ month rest-frame, and can be compared with the total dispersion in the residuals, $\sigma\sim 0.117$. 

\begin{figure}
\centering
\resizebox{\hsize}{!}{\includegraphics{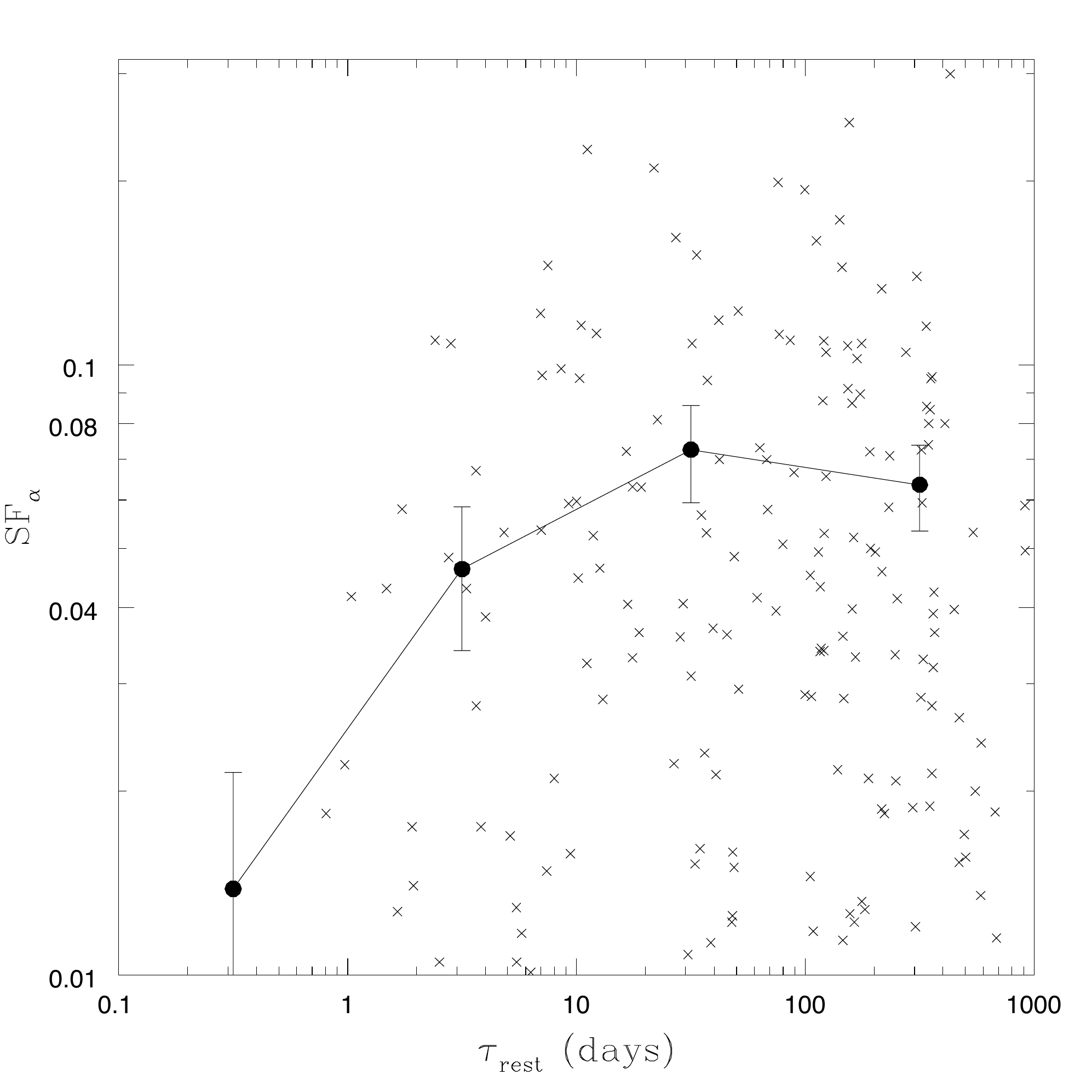}}
\caption{Structure function of $\alpha_{ox}(t)$ vs the rest-frame time lag, for sample B. The crosses represent the variations of individual sources for any couple of epochs. The filled circles connected by continuous lines represent the binned ensemble structure function.}
\end{figure}

As found in Paper I, variability in $\alpha_{ox}$ for individual sources accounts for a large part of the observed dispersion around the average $\alpha_{ox}-L_{UV}$ correlation. We call this ``intra-source dispersion'', while the scatter of the time-average of $\alpha_{ox}$ values for individual sources constitutes the ``inter-source dispersion''. The overall variance is then:

\begin{equation}
\sigma^2=\sigma^2_{\rm intra-source}+\sigma^2_{\rm inter-source}\, .
\end{equation}
Inserting the values 0.073 and 0.117 that we obtained for the intra-source and total dispersions, respectively, Eq. (11) indicates a $\sim 40\%$ contribution of ``intra-source dispersion'' to the total variance $\sigma^2$, similar to Paper I.

\subsection{X-ray SF}

It is also useful to compute separate structure functions for the X-ray and optical variations, to compare the variability properties of this sample with previous analyses. We therefore define:

\begin{equation}
S\negthinspace F_X(\tau)=\sqrt{\pi/2}\,\langle|\log F_X(t+\tau)-\log F_X(t)|\rangle\, ,
\end{equation}
where $F_X$ is the X-ray flux in the observed 0.2-2 keV band. This is similar to the definition introduced by us \citep{vagn11}, except that we omit here the subtraction of the contribution due to photometric noise, which turns out negligible in this case ($\sigma_n\sim 0.01$).

The structure function, shown in Figure 8, represents a variability of $\ga 0.2$ at $\sim$ 1 yr in the logarithm, or $~\sim 60\%$. This can be compared with the SF obtained by us from the {\it XMM-Newton} serendipitous source catalog \citep{vagn11}, which has similar levels of variability at all timescales. The observed X-ray band in that case is 0.5-4.5 keV, which translates to $\sim 1-11$ keV for the higher redshifts of that sample. The luminosities are also different, but the same variability is also found for the lower luminosity sources in the \citet{vagn11} sample, which are comparable to the sources in the present sample. For low-redshift AGNs, most authors use normalised excess variance or fractional variability, and no structure function analyses are available. The energy bands are also usually harder. For example, \citet{mark04} find a fractional variability $F_{var}$ between 10\% and 70\% for 55 Seyfert 1 AGNs at months-years timescale in the 2-4 keV band. \citet{chit09} find an average long-term fractional variability $\sim 60\%$ in the 1.5-3 keV band for $\sim 30$ Seyferts. No direct comparisons are available in the 0.2-2 keV band; energy-dependent analyses \citep{gier06,arev08} report a peak of variability around 1-2 keV for a few Seyferts, with decreasing variability both at lower and higher energies, although different behaviors are found for other sources. With these limitations, our SF in Figure 8 compares reasonably well with other results for similar AGN populations.

\begin{figure}
\centering
\resizebox{\hsize}{!}{\includegraphics{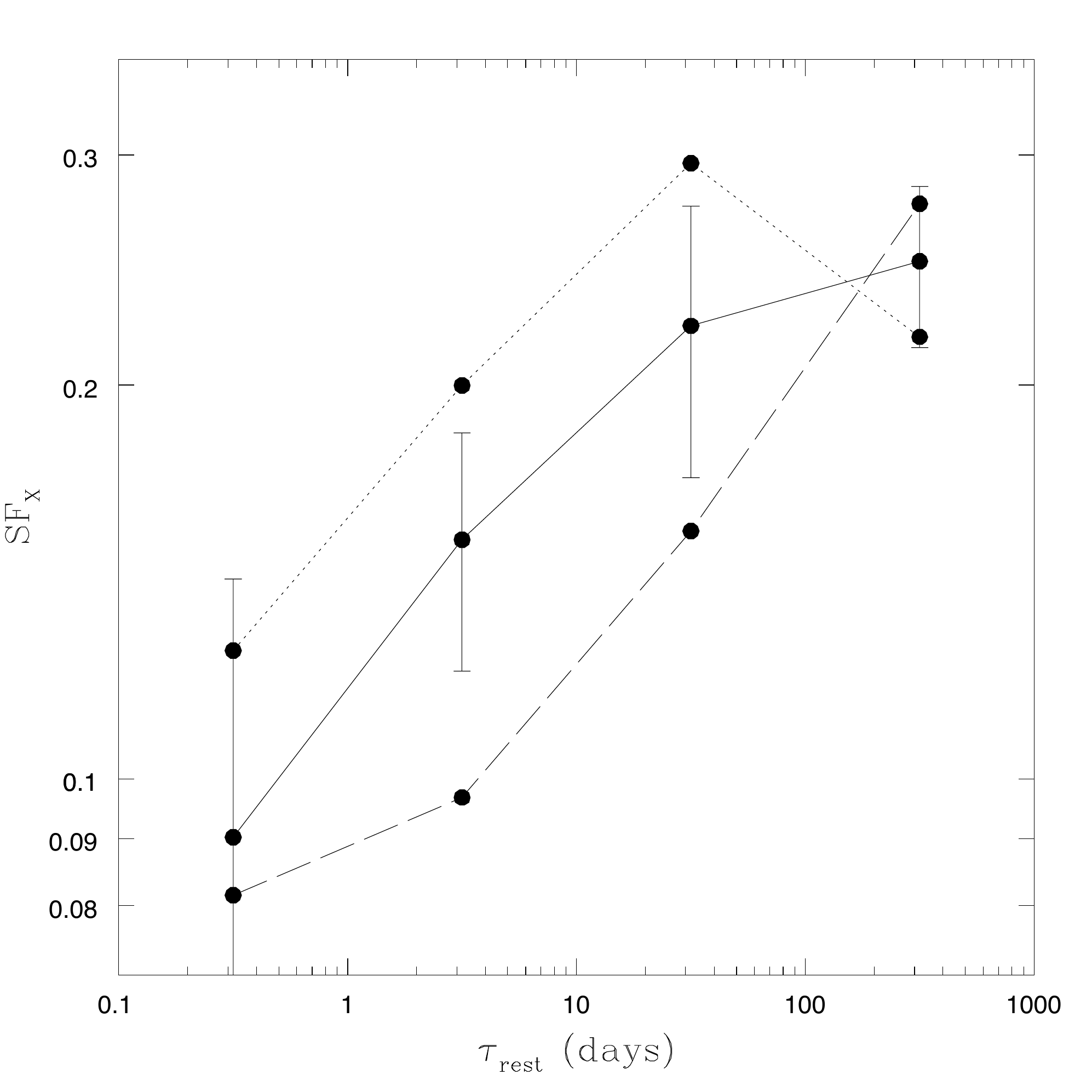}}
\caption{Binned ensemble structure function of the X-ray flux in the 0.2-2 keV band vs the rest-frame time lag. Continuous line: whole sample B; dotted line: NLS1; dashed line: BLS1.}
\end{figure}

\subsection{Optical/UV SF}

Here, we define:

\begin{equation}
S\negthinspace F_O(\tau)=\sqrt{\pi/2}\,\langle|m(t+\tau)-m(t)|\rangle\, ,
\end{equation}
where $m$ is the apparent magnitude in any of the UVOT bands. We omit noise subtraction in this case as well. We stress that optical variabilities measured through Eq. (13) differ from the X-ray variabilities measured through Eq. (12) by the factor 2.5 introduced by magnitudes, which are usually adopted in optical studies.

The result is shown in Figure 9, upper panel, for the 6 UVOT bands, and represents a variability of $\sim 0.3-0.4$ mag at $\sim$ 1 yr.
The values of our SFs can be compared with many other SF analyses, although most of them refer to quasars at higher redshifts. For example, \citet{vand04} find $S\negthinspace F\sim0.2-0.15$ in the SDSS $gri$ bands, and \citet{wilh08} $S\negthinspace F\sim0.3-0.15$ in the $ugriz$ bands, which scan an overall rest-frame range $\sim1400-3600$\AA\ at $\langle z\rangle\sim1.5$. \citet{macl12} find $S\negthinspace F\la 0.2$ in the rest-frame interval 2000-3000\AA. For the low-redshift AGNs of the present sample, the above $\lambda$ intervals are well covered by the 4 harder UVOT bands, $UVW2$, $UVM2$, $UVW1$, $U$, where our SFs are $\ga 0.3$ mag. The variability of our sample is then slightly higher than the variability of the comparison samples. This however is also due to the lower luminosity of our sources, and to the well-known fact that variability decreases with luminosity (e.g., as $L^{-0.246}$ following \citet{vand04}). Therefore, we extract from our sample B a subsample of sources with $\langle L_{UV}\rangle\ge 10^{29}$ erg s$^{-1}$Hz$^{-1}$, well matched with the luminosities of the \citet{vand04} and \citet{wilh08} samples. The SFs for such subsample, shown in the middle panel of Figure 9, amount to $\la 0.2$ mag at 1 yr, similar to the comparison samples.

\begin{figure}
\centering
\resizebox{\hsize}{!}{\includegraphics{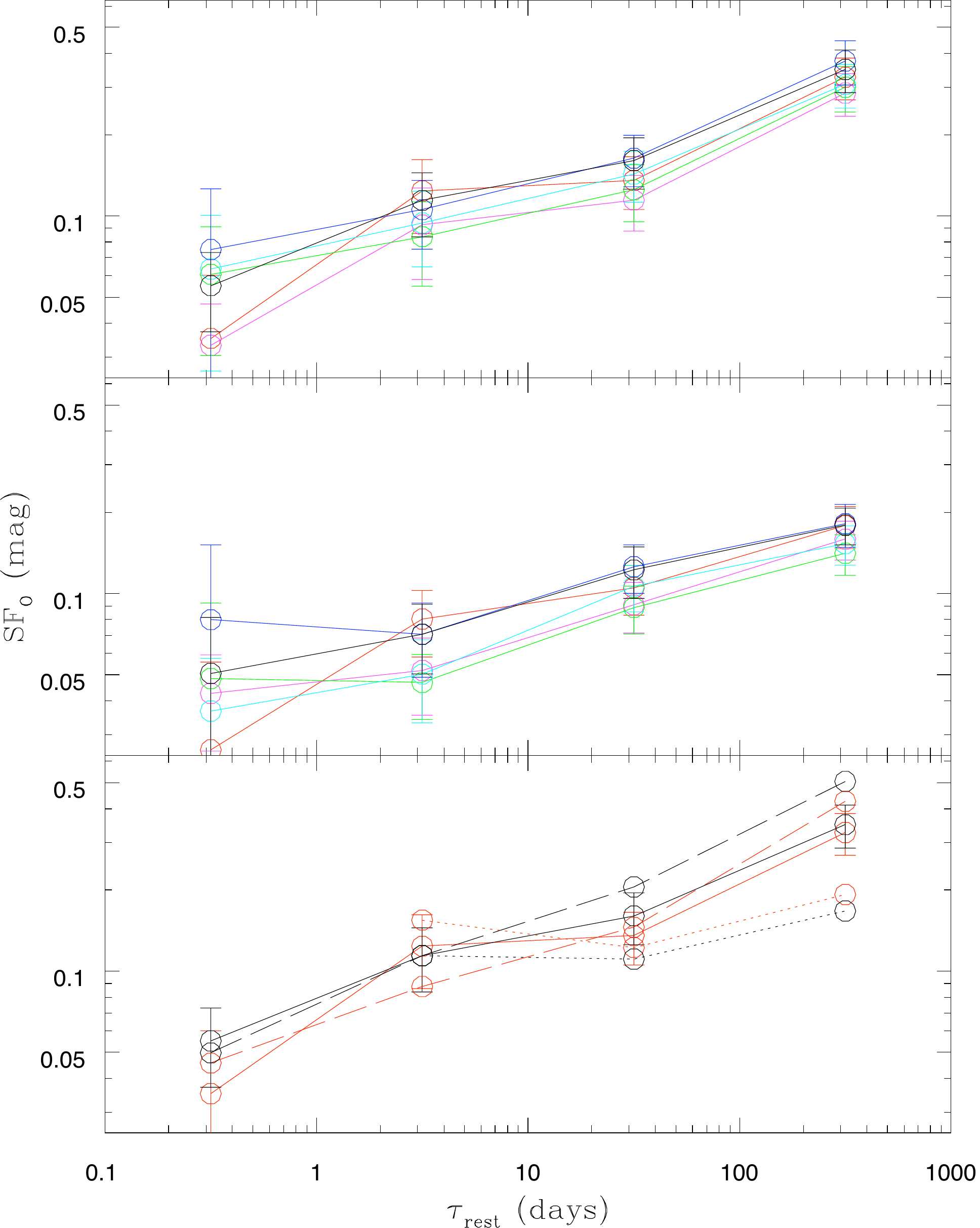}}
\caption{UV/optical structure function. {\it Upper panel}: whole sample B, UVOT filters $UVW2$ (black), $UVM2$ (blue), $UVW1$ (cyan), $U$ (green), $B$ (magenta), $V$ (red). {\it Middle panel}: subsample with $L_{UV}\ge 10^{29}$ erg s$^{-1}$Hz$^{-1}$, same color code. {\it Lower panel}: NLS1 (dotted lines), BLS1 (dashed lines), whole sample B (continuous lines). Only the $UVW2$ filter (black) and the $V$ filter (red) are shown.}
\end{figure}

\subsection{NLS1s}

The presence of several NLS1 AGNs in the sample that we have analysed (28 NLS1 among 61 multi-epoch sources of sample B) gives us the opportunity to measure their variability in comparison with Broad Line Seyfert 1 (BLS1). NLS1 are known to be strongly X-ray variable for timescales $\la 1$ day \citep[e.g.][]{leig99}, and have been suggested to be strongly variable even at longer timescales \citep{hori01}. We show in Figure 8 the X-ray SF of NLS1 (dotted line) and BLS1 (dashed line), compared with the overall behavior of sample B (continuous line). It is seen that NLS1 vary more than the average, and BLS1 less than the average, on timescales shorter than a few months, while there is no such indication for a lag of $\sim 1$ yr. 

We show in the lower panel of Figure 9 the Optical/UV SFs of our NLS1 (dotted lines) and BLS1 (dashed lines), together with the overall sample B (continuous lines), for the extremal UVOT filters $UVW2$ (black) and $V$ (red). In both filters, NLS1 are less variable than the average, and BLS1 more variable. 
Our result confirms previous findings of a weak optical variability of NLS1 \citep[e.g.][]{ai10}.

\section{Discussion}

This paper is the second of a series trying to quantify the contribution of X-ray and UV variability to the dispersion of the $\alpha_{ox}-L_{UV}$ anti-correlation. It is confirmed that this contribution is important ($\sim 40\%$ of the total variance for the sample here analysed), while the ``artificial $\alpha_{ox}$ variability'', present in many analyses because of the non-simultaneity of X-ray and UV/optical observations, turns out to be less important, in the sense that it is surpassed by the ``intrinsic $\alpha_{ox}$ variability''. Indeed, strong X-ray and/or UV changes occur for individual sources: while these variations could in principle occur with minor changes of the X-ray/UV ratio, the strong $\alpha_{ox}$ variations (measured by simultaneous X-ray/UV observations) demonstrate that this is not the case.

Stronger variations occur in the X-rays than in UV. While this behavior could be affected by the presence of many NLS1, strongly variable in X-rays,
we have shown that the average variability properties of the analysed sample do not suggest a special effect of such factor.

We have also discussed the effect of host galaxy dilution on the slope of the $\alpha_{ox}-L_{UV}$ anti-correlation. We have shown that the effect is important for a limited number of low-luminosity sources, still producing a significant flattening of the relation.
Even when corrected for the dilution effect, the $\alpha_{ox}-L_{UV}$ relation remains flatter than those found at high luminosities by \citet{gibs08} and by ourselves (Paper I).

It is interesting to note the recent work
by \citet{sazo12}, who evaluate corona luminosities for a sample of 68 Seyferts through hard X-ray observations by {\it INTEGRAL}, and accretion disk luminosities through {\it Spitzer} observations of the radiation reprocessed by the torus in the mid-infrared, and estimate a disk/corona luminosity ratio approximately constant over 2 decades in luminosity. While this apparently would contradict the $\alpha_{ox}-L_{UV}$ anti-correlation, the authors argue that the 2500 \AA\ luminosity $L_{UV}$ is a good indicator of the accretion disk luminosity for quasars, but not for lower luminosity AGNs, which are expected to have smaller mass black holes, and hotter accretion disks, with emission peaked in the extreme-UV, rather than in the near-UV. This would suggest that $\alpha_{ox}$ is nearly constant at low luminosities, but this indication is not supported by our findings, Eq. (8), nor by those of \citet{xu11}.

The variability of $\alpha_{ox}$, measured by the SFs of Paper I and of the present paper, also gives information on the relation between disk and corona emissions and their variabilities. This relation is complex and includes many processes, e.g variable X-ray irradiation driving optical variations through variable heating of the internal parts of the disk on relatively short timescales, and Compton up-scattering in the corona by UV/optical photons generated in the accretion disk and variable on longer timescales due to disk instabilities born in the outer parts of the disk and propagating inwards \citep[e.g.][]{czer06,arev06}. The SF of $\alpha_{ox}$, shown in Figure 7, increases with the time lag up to $\sim 1$ month, while in Paper I a further increase at $\sim 1$ year was also present. These findings indicate the important contribution of the intermediate-long timescale variations, possibly generated in the outer parts of the accretion disk.

Another interesting issue concerns the ``inter-source dispersion'', i.e. the residual dispersion of the $\alpha_{ox}-L_{UV}$ relation after accounting for the effect of variability. This could be related to the dependence of $\alpha_{ox}$ on a second physical parameter, besides the primary dependence on luminosity. For example, both \citet{luss10} and G10 find evidence of a decrease of $\alpha_{ox}$ with Eddington ratio. Moreover, \citet{youn10} find significant partial anti-correlation with the Eddington ratio, when dependence on $L_{UV}$ is accounted for; the significance increases if the X-ray energy in the $\alpha_{ox}$ definition is moved upwards.

A further step in the investigation of the variability of $\alpha_{ox}$ would be a better temporal sampling of the simultaneous X-ray and optical observations. An appropriate strategy would be a multi-epoch survey of the same field with an X-ray/optical telescope such as {\it XMM-Newton} or {\it Swift}, and the opportunity is punctually offered by the XMM Deep survey in the Chandra Deep Field South \citep{coma11}. We are preparing an analysis of the individual properties of $\alpha_{ox}$ variability for the brightest sources, which have simultaneous X-ray/optical multi-epoch information (Vagnetti et al, in preparation).

\begin{acknowledgements}
We thank Dirk Grupe for help in the use of his published data, and for suggestions and comments. We thank Maurizio Paolillo, Sara Turriziani, and Guido Risaliti for useful discussions. We thank the referee, who induced us to correct for the host galaxy contamination. F.V. acknowledges hospitality by University of Rome ``La Sapienza", during his sabbatical leave from University of Rome ``Tor Vergata". This research has made use of data and/or software provided by the High Energy Astrophysics Science Archive Research Center (HEASARC), which is a service of the Astrophysics Science Division at NASA/GSFC and the High Energy Astrophysics Division of the Smithsonian Astrophysical Observatory. This research has made use of the NASA/IPAC Extragalactic Database (NED), which is operated by the Jet Propulsion Laboratory, California Institute of Technology, under contract with the National Aeronautics and Space Administration. No financial resources have been granted to this research by the University of Roma Tor Vergata.
\end{acknowledgements}

\bibliographystyle{aa}
\bibliography{alox2.bib}{}

\onltab{1}{
\begin{longtable}{r l c c l l l l l l r}
\caption{The sources.}\\
\hline
$N_{sou}$&name&$N_{epo}$&epoch (MJD)&$z$&$\alpha_x$&$\log L_X$&$\log L_{UV}$&$\log L_G$&$\alpha_{ox}$&$f_{RL}$\\
\hline
(1)&(2)&(3)&(4)&(5)&(6)&(7)&(8)&(9)&(10)&(11)\\
\hline
\endfirsthead
\caption{continued.}\\
\hline
$N_{sou}$&name&$N_{epo}$&epoch (MJD)&$z$&$\alpha_x$&$\log L_X$&$\log L_{UV}$&$\log L_G$&$\alpha_{ox}$&$f_{RL}$\\
\hline
\endhead
\hline
\endfoot
  1 & Mkn\_335 & 1 & 54632.1 & 0.026 & 1.31 & 25.61 & 28.85 & 27.68 & -1.24 & 0\\
  2 & ESO 242-G008 & 1 & 53964.1 & 0.059 & 0.74 & 25.59 & 28.3 & 27.98 & -1.04 & -1\\
   &  & 2 & 54044.6 &  & 1.17 & 25.38 & 28.61 &  & -1.24 & \\
  3 & Ton S 180 & 1 & 53990.3 & 0.062 & 1.56 & 25.89 & 29.65 & 27.77 & -1.44 & -1\\
   &  & 2 & 54236.4 &  & 1.48 & 26.0 & 29.6 &  & -1.38 & \\
   &  & 3 & 54465.4 &  & 1.58 & 26.08 & 29.73 &  & -1.4 & \\
  4 & QSO 0056-36 & 1 & 53873.9 & 0.165 & 1.06 & 26.58 & 30.18 & $-$ & -1.38 & -1\\
   &  & 2 & 53878.6 &  & 1.13 & 26.46 & 30.15 &  & -1.42 & \\
  5 & RX J0100.4-5113 & 1 & 54454.5 & 0.062 & 1.08 & 25.54 & 29.12 & 28.03 & -1.37 & -1\\
   &  & 2 & 54722.1 &  & 1.19 & 25.43 & 29.11 &  & -1.41 & \\
   &  & 3 & 54776.5 &  & 1.36 & 25.58 & 29.18 &  & -1.38 & \\
   &  & 4 & 54794.2 &  & 1.36 & 25.71 & 29.21 &  & -1.34 & \\
  6 & RX J0105.6-1416 & 1 & 54475.7 & 0.07 & 0.92 & 25.92 & 29.14 & 28.14 & -1.23 & -1\\
   &  & 2 & 54599.9 &  & 0.99 & 25.96 & 29.26 &  & -1.27 & \\
   &  & 3 & 54633.4 &  & 1.14 & 25.98 & 29.27 &  & -1.26 & \\
   &  & 4 & 54643.5 &  & 1.02 & 25.97 & 29.22 &  & -1.25 & \\
  7 & RX J0117.5-3826 & 1 & 54646.7 & 0.225 & 1.73 & 26.1 & 29.58 & 28.78 & -1.34 & -1\\
   &  & 2 & 54949.5 &  & 2.14 & 26.03 & 29.51 &  & -1.33 & \\
   &  & 3 & 55145.5 &  & 2.09 & 25.79 & 29.49 &  & -1.42 & \\
  8 & MS 0117-28 & 1 & 54044.3 & 0.349 & 1.6 & 26.37 & 30.53 & $-$ & -1.6 & -1\\
  9 & RX J0128.1-1848 & 1 & 54275.5 & 0.046 & 0.92 & 25.76 & 28.89 & 28.23 & -1.2 & -1\\
   &  & 2 & 54501.5 &  & 0.89 & 25.69 & 28.83 &  & -1.21 & \\
   &  & 3 & 54502.5 &  & 1.1 & 25.71 & 28.84 &  & -1.2 & \\
   &  & 4 & 54503.5 &  & 1.01 & 25.74 & 28.84 &  & -1.19 & \\
  10 & RX J0134.2-4258 & 1 & 54433.5 & 0.237 & 1.29 & 26.26 & 30.25 & $-$ & -1.53 & 1\\
  11 & RX J0136.9-3510 & 1 & 53767.0 & 0.289 & 1.87 & 26.54 & 29.79 & 28.29 & -1.25 & -1\\
  12 & RX J0148.3-2758 & 1 & 54037.3 & 0.121 & 1.75 & 26.33 & 29.71 & 28.68 & -1.3 & -1\\
   &  & 2 & 54232.2 &  & 1.82 & 26.12 & 29.72 &  & -1.38 & \\
   &  & 3 & 54429.6 &  & 1.73 & 26.35 & 29.68 &  & -1.28 & \\
   &  & 4 & 54593.2 &  & 1.81 & 26.25 & 29.67 &  & -1.31 & \\
   &  & 5 & 54633.9 &  & 1.67 & 26.34 & 29.7 &  & -1.29 & \\
   &  & 6 & 54644.3 &  & 1.57 & 26.13 & 29.65 &  & -1.35 & \\
  13 & RX J0152.4-2319 & 1 & 54467.7 & 0.113 & 1.25 & 25.99 & 29.69 & 28.33 & -1.42 & 0\\
   &  & 2 & 54508.9 &  & 1.24 & 26.14 & 29.7 &  & -1.37 & \\
   &  & 3 & 54514.6 &  & 1.23 & 26.2 & 29.71 &  & -1.35 & \\
  14 & Mkn 1044 & 1 & 54306.6 & 0.017 & 1.46 & 25.01 & 28.33 & 27.35 & -1.28 & 1\\
   &  & 2 & 54313.5 &  & 1.47 & 24.68 & 28.34 &  & -1.41 & \\
  15 & Mkn 1048 & 1 & 54304.5 & 0.042 & 0.61 & 25.42 & 29.13 & 27.87 & -1.42 & 0\\
   &  & 2 & 54451.8 &  & 1.02 & 25.94 & 29.2 &  & -1.25 & \\
   &  & 3 & 54529.4 &  & 0.71 & 25.82 & 29.18 &  & -1.29 & \\
   &  & 4 & 54624.3 &  & 0.78 & 25.84 & 29.18 &  & -1.28 & \\
  16 & RX J0311.3-2046 & 1 & 54916.7 & 0.07 & 0.83 & 25.93 & 28.99 & 28.0 & -1.18 & -1\\
   &  & 2 & 54982.6 &  & 0.87 & 25.85 & 29.02 &  & -1.22 & \\
  17 & RX J0319.8-2627 & 1 & 54166.5 & 0.076 & 0.82 & 25.6 & 28.71 & 28.27 & -1.2 & -1\\
   &  & 2 & 54169.2 &  & 0.94 & 25.49 & 28.63 &  & -1.21 & \\
   &  & 3 & 54174.5 &  & 0.87 & 25.54 & 28.69 &  & -1.21 & \\
   &  & 4 & 54175.1 &  & 0.7 & 25.51 & 28.68 &  & -1.22 & \\
   &  & 5 & 54546.8 &  & 1.08 & 25.62 & 28.98 &  & -1.29 & \\
   &  & 6 & 54559.5 &  & 1.05 & 25.83 & 29.06 &  & -1.24 & \\
   &  & 7 & 54798.3 &  & 0.55 & 25.36 & 28.54 &  & -1.22 & \\
   &  & 8 & 54911.5 &  & 0.98 & 25.59 & 28.73 &  & -1.2 & \\
  18 & RX J0323.2-4911 & 1 & 54441.9 & 0.071 & 1.33 & 25.01 & 27.77 & 28.23 & -1.06 & -1\\
   &  & 2 & 54449.4 &  & 1.08 & 25.29 & 27.73 &  & -0.94 & \\
  19 & ESO 301-G13 & 1 & 54779.6 & 0.059 & 1.35 & 25.62 & 28.96 & 28.01 & -1.28 & -1\\
   &  & 2 & 54810.6 &  & 1.19 & 25.57 & 28.81 &  & -1.24 & \\
   &  & 3 & 55122.4 &  & 1.1 & 25.75 & 28.93 &  & -1.22 & \\
   &  & 4 & 55194.8 &  & 1.31 & 25.6 & 28.94 &  & -1.28 & \\
  20 & VCV 0331-37 & 1 & 54650.7 & 0.064 & 1.04 & 25.53 & 28.81 & 27.62 & -1.26 & -1\\
   &  & 2 & 54766.1 &  & 1.07 & 25.5 & 28.8 &  & -1.27 & \\
   &  & 3 & 54913.6 &  & 1.24 & 25.51 & 28.87 &  & -1.29 & \\
   &  & 4 & 54915.3 &  & 1.19 & 25.55 & 28.88 &  & -1.28 & \\
  21 & RX J0349.1-4711 & 1 & 53974.5 & 0.299 & 1.62 & 26.39 & 29.98 & 28.65 & -1.38 & -1\\
   &  & 2 & 54185.9 &  & 1.35 & 26.53 & 29.98 &  & -1.33 & \\
  22 & Fairall 1116 & 1 & 53752.1 & 0.059 & 1.05 & 25.94 & 29.17 & 28.13 & -1.24 & -1\\
   &  & 2 & 53833.6 &  & 1.33 & 25.74 & 29.26 &  & -1.35 & \\
   &  & 3 & 53857.5 &  & 1.35 & 25.54 & 29.28 &  & -1.43 & \\
  23 & Fairall 1119 & 1 & 54751.9 & 0.055 & 0.72 & 25.39 & 27.65 & 28.07 & -0.87 & -1\\
   &  & 2 & 54765.2 &  & 0.6 & 25.3 & 27.69 &  & -0.92 & \\
  24 & RX J0412.7-4712 & 1 & 54459.3 & 0.132 & 1.02 & 26.42 & 29.68 & 28.57 & -1.25 & -1\\
   &  & 2 & 54467.3 &  & 1.09 & 26.61 & 29.72 &  & -1.2 & \\
  25 & 1H 0419-577 & 1 & 54761.4 & 0.104 & 1.22 & 26.81 & 29.92 & 28.9 & -1.19 & 0\\
   &  & 2 & 54782.6 &  & 1.07 & 26.62 & 29.9 &  & -1.26 & \\
  26 & Fairall 303 & 1 & 54768.2 & 0.04 & 1.32 & 25.23 & 28.37 & 27.47 & -1.2 & -1\\
   &  & 2 & 54787.7 &  & 1.17 & 25.08 & 28.31 &  & -1.24 & \\
  27 & RX J0437.4-4711 & 1 & 54441.6 & 0.052 & 1.2 & 25.67 & 29.0 & 28.15 & -1.28 & -1\\
   &  & 2 & 54452.1 &  & 1.17 & 25.83 & 29.0 &  & -1.22 & \\
  28 & RX J0439.6-5311 & 1 & 53741.3 & 0.243 & 2.16 & 26.67 & 29.7 & $-$ & -1.17 & -1\\
   &  & 2 & 53838.3 &  & 2.12 & 26.64 & 29.7 &  & -1.17 & \\
   &  & 3 & 53840.5 &  & 2.05 & 26.79 & 29.7 &  & -1.11 & \\
   &  & 4 & 53873.7 &  & 2.07 & 26.73 & 29.69 &  & -1.14 & \\
  29 & RX J0859.0+4866 & 1 & 54020.5 & 0.083 & 0.91 & 25.94 & 29.23 & 28.4 & -1.26 & -1\\
   &  & 2 & 54225.1 &  & 0.98 & 25.92 & 29.26 &  & -1.28 & \\
   &  & 3 & 54239.3 &  & 1.14 & 25.91 & 29.32 &  & -1.31 & \\
  30 & RX J0902.5-0700 & 1 & 54269.1 & 0.089 & 1.23 & 25.16 & 28.63 & 27.68 & -1.33 & -1\\
   &  & 2 & 54461.6 &  & 1.24 & 25.34 & 28.77 &  & -1.32 & \\
   &  & 3 & 54467.5 &  & 1.61 & 25.31 & 28.78 &  & -1.33 & \\
  31 & Mkn 110 & 1 & 55202.8 & 0.035 & 0.98 & 26.13 & 29.07 & 26.4 & -1.13 & 0\\
   &  & 2 & 55208.7 &  & 1.02 & 26.12 & 29.09 &  & -1.14 & \\
  33 & RX J1005.7+4332 & 1 & 53789.6 & 0.178 & 1.8 & 25.99 & 29.84 & $-$ & -1.48 & 0\\
  34 & RX J1007.1+2203 & 1 & 54281.8 & 0.083 & 1.5 & 25.4 & 28.71 & 27.76 & -1.27 & -1\\
   &  & 2 & 54647.4 &  & 1.53 & 25.11 & 28.72 &  & -1.39 & \\
  35 & CBS 126 & 1 & 53899.2 & 0.079 & 1.4 & 25.6 & 29.3 & 28.21 & -1.42 & -1\\
   &  & 2 & 54132.7 &  & 1.39 & 25.78 & 29.37 &  & -1.38 & \\
  36 & Mkn 141 & 1 & 54023.5 & 0.042 & 0.76 & 25.08 & 28.17 & 28.05 & -1.19 & 0\\
   &  & 2 & 54185.9 &  & 0.36 & 24.52 & 28.26 &  & -1.44 & \\
  37 & Mkn 142 & 1 & 54428.7 & 0.045 & 1.38 & 25.17 & 28.63 & 27.61 & -1.33 & 0\\
   &  & 2 & 54479.9 &  & 1.72 & 25.41 & 28.74 &  & -1.28 & \\
  38 & RX J1117.1+6522 & 1 & 54054.5 & 0.147 & 1.93 & 25.78 & 29.36 & 28.87 & -1.37 & -1\\
   &  & 2 & 54182.8 &  & 1.05 & 25.26 & 29.25 &  & -1.53 & \\
   &  & 3 & 54195.6 &  & 2.09 & 25.98 & 29.39 &  & -1.31 & \\
   &  & 4 & 54220.7 &  & 1.91 & 25.4 & 29.36 &  & -1.52 & \\
  39 & Ton 1388 & 1 & 54289.6 & 0.177 & 1.26 & 26.71 & 30.64 & $-$ & -1.51 & 0\\
   &  & 2 & 54940.4 &  & 1.43 & 26.7 & 30.67 &  & -1.53 & \\
  40 & EXO 1128+6908 & 1 & 54878.8 & 0.045 & 1.24 & 25.58 & 28.6 & 27.93 & -1.19 & -1\\
  41 & B2 1128+31 & 1 & 54753.7 & 0.289 & 1.05 & 26.89 & 30.42 & $-$ & -1.35 & 1\\
   &  & 2 & 54879.7 &  & 0.99 & 26.96 & 30.35 &  & -1.3 & \\
   &  & 3 & 55163.7 &  & 1.14 & 26.82 & 30.32 &  & -1.34 & \\
  42 & SBS 1136+579 & 1 & 54182.5 & 0.116 & 0.98 & 25.83 & 29.14 & 28.26 & -1.27 & -1\\
   &  & 2 & 54936.6 &  & 1.32 & 25.35 & 28.71 &  & -1.29 & \\
  43 & CASG 855 & 1 & 54883.7 & 0.04 & 0.83 & 25.0 & 27.21 & 27.76 & -0.85 & -1\\
  44 & NGC 4051 & 1 & 54876.1 & 0.0020 & 1.59 & 23.72 & 26.35 & 26.08 & -1.01 & 0\\
  45 & GQ Comae & 1 & 54762.5 & 0.165 & 1.1 & 26.67 & 29.82 & $-$ & -1.21 & -1\\
   &  & 2 & 54763.5 &  & 1.01 & 26.7 & 29.8 &  & -1.19 & \\
  46 & RX J1209.8+3217 & 1 & 54269.0 & 0.145 & 1.86 & 25.26 & 29.08 & 28.26 & -1.47 & 0\\
   &  & 2 & 54760.5 &  & 3.07 & 24.6 & 29.2 &  & -1.76 & \\
  47 & PG 1211+143 & 1 & 54192.9 & 0.082 & 1.89 & 25.28 & 29.81 & 28.74 & -1.74 & 0\\
  48 & Mkn 766 & 1 & 54090.9 & 0.013 & 1.05 & 24.73 & 26.33 & 27.45 & -0.61 & 1\\
  49 & 3C 273 & 1 & 54863.9 & 0.158 & 0.72 & 27.82 & 31.19 & $-$ & -1.29 & 0\\
   &  & 2 & 54919.5 &  & 0.74 & 27.79 & 31.19 &  & -1.3 & \\
  50 & RX J1231.6+7044 & 1 & 54236.0 & 0.208 & 0.6 & 26.88 & 29.96 & 28.54 & -1.18 & -1\\
   &  & 2 & 54282.7 &  & 0.82 & 26.86 & 29.97 &  & -1.19 & \\
   &  & 3 & 54293.6 &  & 0.85 & 26.83 & 29.94 &  & -1.19 & \\
   &  & 4 & 54294.2 &  & 0.87 & 26.82 & 29.94 &  & -1.2 & \\
  51 & MCG+08-23-006 & 1 & 54190.0 & 0.03 & 0.82 & 24.52 & 27.31 & 27.62 & -1.07 & 0\\
  52 & NGC 4593 & 1 & 54693.8 & 0.0090 & 0.69 & 24.64 & 27.22 & 27.25 & -0.99 & 0\\
  53 & RX J1304.2+0205 & 1 & 54444.5 & 0.229 & 1.97 & 25.97 & 29.64 & 27.84 & -1.41 & -1\\
   &  & 2 & 54681.5 &  & 2.17 & 26.16 & 29.7 &  & -1.36 & \\
  54 & PG 1307+085 & 1 & 54696.0 & 0.155 & 1.16 & 26.44 & 30.09 & $-$ & -1.4 & 0\\
   &  & 2 & 54697.4 &  & 1.26 & 26.44 & 30.09 &  & -1.4 & \\
  55 & RX J1319.9+5235 & 1 & 54028.3 & 0.092 & 1.87 & 25.37 & 28.07 & 27.81 & -1.04 & -1\\
   &  & 2 & 54187.9 &  & 1.86 & 25.45 & 28.12 &  & -1.03 & \\
   &  & 3 & 54195.6 &  & 1.64 & 25.65 & 28.07 &  & -0.93 & \\
  57 & Ton 730 & 1 & 55057.2 & 0.087 & 1.41 & 25.83 & 29.3 & 27.02 & -1.33 & -1\\
  58 & RX J1355.2+5612 & 1 & 54233.5 & 0.122 & 1.77 & 25.77 & 29.14 & 28.41 & -1.3 & 0\\
   &  & 2 & 54241.9 &  & 1.66 & 26.12 & 29.11 &  & -1.15 & \\
   &  & 3 & 54272.4 &  & 1.61 & 25.69 & 29.1 &  & -1.31 & \\
   &  & 4 & 54277.8 &  & 1.79 & 25.84 & 29.12 &  & -1.26 & \\
   &  & 5 & 54279.5 &  & 2.09 & 25.7 & 29.09 &  & -1.3 & \\
  59 & PG 1402+261 & 1 & 53923.3 & 0.164 & 1.35 & 26.35 & 30.19 & $-$ & -1.48 & -1\\
   &  & 2 & 53948.4 &  & 1.46 & 26.35 & 30.2 &  & -1.48 & \\
   &  & 3 & 53960.9 &  & 1.39 & 26.39 & 30.22 &  & -1.47 & \\
  60 & RX J1413.6+7029 & 1 & 54600.8 & 0.107 & 0.64 & 25.69 & 28.14 & 28.24 & -0.94 & -1\\
   &  & 2 & 54905.7 &  & 0.97 & 26.02 & 28.74 &  & -1.04 & \\
  61 & NGC 5548 & 1 & 54270.4 & 0.017 & 0.41 & 24.65 & 27.67 & 27.59 & -1.16 & 0\\
  62 & Mkn 813 & 1 & 54110.3 & 0.111 & 1.0 & 26.38 & 30.03 & 27.05 & -1.4 & -1\\
  63 & Mkn 684 & 1 & 53879.5 & 0.046 & 1.42 & 25.27 & 29.07 & 28.1 & -1.46 & 1\\
   &  & 2 & 53880.5 &  & 1.24 & 25.42 & 29.1 &  & -1.41 & \\
  64 & Mkn 478 & 1 & 53976.8 & 0.077 & 1.41 & 25.52 & 29.71 & 28.35 & -1.61 & 0\\
   &  & 2 & 54017.0 &  & 1.35 & 25.76 & 29.71 &  & -1.51 & \\
  66 & Mkn 841 & 1 & 54101.5 & 0.036 & 0.87 & 25.54 & 28.97 & 27.59 & -1.32 & 0\\
   &  & 2 & 54133.3 &  & 0.96 & 25.59 & 29.0 &  & -1.31 & \\
   &  & 3 & 54134.4 &  & 0.91 & 25.49 & 29.0 &  & -1.35 & \\
   &  & 4 & 54621.3 &  & 1.05 & 25.59 & 29.06 &  & -1.33 & \\
  67 & Mkn 493 & 1 & 53679.7 & 0.032 & 1.29 & 25.01 & 28.57 & 27.65 & -1.37 & 0\\
   &  & 2 & 53682.6 &  & 1.32 & 25.12 & 28.4 &  & -1.26 & \\
   &  & 3 & 54622.5 &  & 1.12 & 25.0 & 28.43 &  & -1.32 & \\
  68 & Mkn 876 & 1 & 53885.7 & 0.129 & 0.9 & 26.42 & 30.26 & 28.4 & -1.47 & 0\\
   &  & 2 & 53905.5 &  & 1.01 & 26.35 & 30.28 &  & -1.51 & \\
  70 & KUG 1618+410 & 1 & 54475.9 & 0.038 & 0.9 & 24.63 & 27.75 & 27.63 & -1.2 & -1\\
   &  & 2 & 54479.7 &  & 0.98 & 24.67 & 27.61 &  & -1.13 & \\
  71 & PG 1626+554 & 1 & 54606.4 & 0.133 & 1.11 & 26.33 & 29.82 & $-$ & -1.34 & -1\\
   &  & 2 & 54618.9 &  & 1.32 & 26.27 & 29.84 &  & -1.37 & \\
  72 & EXO 1627+40 & 1 & 54277.3 & 0.272 & 1.15 & 26.56 & 29.55 & 27.9 & -1.15 & 1\\
   &  & 2 & 54473.3 &  & 1.06 & 26.61 & 29.53 &  & -1.12 & \\
   &  & 3 & 54477.1 &  & 1.41 & 26.4 & 29.55 &  & -1.21 & \\
  73 & RX J1702.5+3247 & 1 & 53978.5 & 0.164 & 1.73 & 26.22 & 29.94 & 28.94 & -1.43 & 0\\
   &  & 2 & 54019.5 &  & 1.79 & 26.37 & 29.95 &  & -1.37 & \\
   &  & 3 & 54119.5 &  & 1.16 & 26.08 & 29.94 &  & -1.48 & \\
   &  & 4 & 54123.4 &  & 1.67 & 26.19 & 29.94 &  & -1.44 & \\
  74 & II Zw 136 & 1 & 54665.5 & 0.065 & 1.32 & 26.07 & 29.65 & 28.31 & -1.38 & 0\\
   &  & 2 & 54666.6 &  & 1.38 & 26.09 & 29.65 &  & -1.37 & \\
   &  & 3 & 54684.2 &  & 1.49 & 25.82 & 29.57 &  & -1.44 & \\
   &  & 4 & 54790.5 &  & 1.24 & 25.92 & 29.59 &  & -1.41 & \\
  76 & RX J2216.8-4451 & 1 & 54659.4 & 0.136 & 1.5 & 26.09 & 29.8 & $-$ & -1.43 & -1\\
   &  & 2 & 54661.6 &  & 1.69 & 26.13 & 29.8 &  & -1.41 & \\
   &  & 3 & 54671.3 &  & 1.66 & 26.4 & 29.81 &  & -1.31 & \\
   &  & 4 & 54673.3 &  & 1.48 & 26.41 & 29.83 &  & -1.31 & \\
  77 & RX J2217.9-5941 & 1 & 54042.1 & 0.16 & 2.01 & 25.45 & 29.44 & 28.3 & -1.53 & 0\\
   &  & 2 & 54220.1 &  & 2.45 & 25.21 & 29.44 &  & -1.62 & \\
  78 & RX J2242.6-3845 & 1 & 54665.9 & 0.221 & 1.68 & 26.04 & 29.65 & 28.18 & -1.38 & -1\\
   &  & 2 & 54668.9 &  & 1.6 & 26.36 & 29.68 &  & -1.27 & \\
  79 & RX J2245.3-4652 & 1 & 53898.5 & 0.201 & 1.16 & 26.28 & 30.23 & 29.28 & -1.52 & -1\\
   &  & 2 & 53901.8 &  & 1.27 & 26.16 & 30.24 &  & -1.56 & \\
  81 & MS 2254-36 & 1 & 54274.9 & 0.039 & 1.19 & 25.22 & 28.3 & 27.73 & -1.18 & -1\\
   &  & 2 & 54316.6 &  & 1.14 & 25.24 & 28.34 &  & -1.19 & \\
   &  & 3 & 54441.5 &  & 1.14 & 25.08 & 28.27 &  & -1.22 & \\
   &  & 4 & 54445.3 &  & 1.18 & 25.19 & 28.31 &  & -1.2 & \\
  82 & RX J2258.7-2609 & 1 & 54671.1 & 0.076 & 0.71 & 25.85 & 28.76 & 28.01 & -1.12 & -1\\
   &  & 2 & 54680.8 &  & 0.98 & 25.88 & 28.8 &  & -1.12 & \\
   &  & 3 & 54725.8 &  & 0.88 & 25.66 & 28.89 &  & -1.24 & \\
   &  & 4 & 54794.0 &  & 0.91 & 25.7 & 28.74 &  & -1.17 & \\
  83 & RX J2301.6-5913 & 1 & 54673.3 & 0.149 & 0.82 & 26.48 & 29.31 & 28.34 & -1.09 & 1\\
   &  & 2 & 54681.2 &  & 0.84 & 26.5 & 29.33 &  & -1.09 & \\
   &  & 3 & 55032.7 &  & 0.77 & 26.62 & 29.3 &  & -1.03 & \\
   &  & 4 & 55098.8 &  & 0.86 & 26.57 & 29.42 &  & -1.09 & \\
   &  & 5 & 55102.8 &  & 0.83 & 26.61 & 29.42 &  & -1.08 & \\
  84 & RX J2301.8-5508 & 1 & 53700.6 & 0.14 & 1.49 & 25.8 & 29.65 & 28.98 & -1.48 & -1\\
   &  & 2 & 53712.2 &  & 1.38 & 25.93 & 29.67 &  & -1.44 & \\
  85 & RX J2304.6-3501 & 1 & 53997.0 & 0.042 & 1.06 & 24.72 & 27.8 & 27.49 & -1.18 & -1\\
  86 & RX J2312.5-3404 & 1 & 54320.5 & 0.202 & 0.67 & 26.23 & 29.76 & 28.64 & -1.35 & 1\\
   &  & 2 & 54457.7 &  & 0.77 & 26.42 & 29.72 &  & -1.27 & \\
  87 & RX J2317.8-4422 & 1 & 53843.5 & 0.132 & 2.58 & 25.2 & 29.14 & 28.3 & -1.51 & -1\\
  88 & RX J2325.2-3236 & 1 & 54002.1 & 0.216 & 1.08 & 26.68 & 29.97 & $-$ & -1.26 & 1\\
   &  & 2 & 54045.8 &  & 1.02 & 26.64 & 29.99 &  & -1.28 & \\
   &  & 3 & 54708.0 &  & 1.21 & 26.6 & 29.92 &  & -1.27 & \\
  89 & IRAS23226-3843 & 1 & 54676.3 & 0.036 & 0.6 & 25.35 & 28.18 & 28.15 & -1.08 & 0\\
  90 & MS 23409-1511 & 1 & 54430.4 & 0.137 & 1.79 & 26.1 & 29.71 & 28.61 & -1.39 & -1\\
   &  & 2 & 54481.5 &  & 1.77 & 26.15 & 29.75 &  & -1.38 & \\
   &  & 3 & 54484.8 &  & 1.87 & 26.13 & 29.72 &  & -1.38 & \\
   &  & 4 & 54485.9 &  & 1.78 & 26.09 & 29.74 &  & -1.4 & \\
   &  & 5 & 54621.6 &  & 1.91 & 25.86 & 29.74 &  & -1.49 & \\
  91 & RX J2349.4-3126 & 1 & 54231.9 & 0.135 & 1.03 & 25.87 & 28.84 & 28.6 & -1.14 & -1\\
   &  & 2 & 54792.8 &  & 1.03 & 25.88 & 28.77 &  & -1.11 & \\
   &  & 3 & 55098.5 &  & 1.03 & 25.88 & 28.66 &  & -1.07 & \\
  92 & AM 2354-304 & 1 & 54850.5 & 0.033 & 1.16 & 25.01 & 28.42 & 27.79 & -1.31 & -1\\
   &  & 2 & 54994.8 &  & 1.21 & 24.9 & 28.42 &  & -1.35 & \\
\end{longtable}
}

\end{document}